\documentclass[]{pasj01}
\usepackage{pict2e}
\usepackage{datetime}
\usepackage{url}
\usepackage[switch,mathlines]{lineno}
\makeatletter
\DeclareRobustCommand{\sector}{\mathord{\mathpalette\make@sector\relax}}
\newcommand{\make@sector}[2]{%
  \settoheight{\unitlength}{$#1x$}%
  \begin{picture}(1,1.06)
  \linethickness{.08\unitlength}
  \moveto(0.5,0)
  \lineto(0.842,1)
  \curveto(.6,1.08)(.4,1.08)(0.158,1)
  \closepath
  \strokepath
  \end{picture}%
}
\makeatother

\begin{document} 

\title{Spin Parity of Spiral Galaxies IV - Differential Reddening of Globular Cluster Systems of Nearby Spiral Galaxies
 }

\author{Masanori \textsc{Iye}\altaffilmark{1,2}}
\altaffiltext{1}{National Astronomical Observatory of Japan, Osawa 2-21-1, Mitaka, Tokyo 181-8588 Japan}
\altaffiltext{2}{The Japan Academy,7-32 Ueno Park, Taito-ku, Tokyo110-0007, Japan}
\email{m.iye@nao.ac.jp}
\author{Masafumi \textsc{Yagi}\altaffilmark{1}}


\maketitle
\nolinenumbers

\begin{abstract}
The northwest side of the disk of M31 is known to be the near side because of the differential reddening of globular clusters found from their photographic photometry. This paper reports a simple geometric model to evaluate the visibility of the effect and its application to published CCD photometry on globular cluster systems of three spiral galaxies, M31, M33, and NGC253. The color difference of globular cluster systems due to differential reddening was confirmed for M31 and NGC253; however, the data for M33 were insufficient.  
The analysis reaffirms the currently adopted interpretation that the side on the minor axis of the galactic disk, where more conspicuous dust features and interstellar reddening are visible, is the nearer side to us and provides an additional basis for using spiral galaxies with identified spiral windings, S-wise or Z-wise, to study the large-scale spin distribution of galaxies in the universe.
\end{abstract}
%\pagewiselinenumbers
\nolinenumbers

\section{Introduction}
Identifying which side on the minor axis of the tilted galactic disk as projected on the sky plane is the nearer side to us has not been a straightforward question from an observational point of view. Indeed, back in the 1950s, there was a big debate between de Vaucouleurs and Lindblad. The former interpreted dust lanes on the near side of the disk as obscuring the central bulge, whereas the latter interpreted that the dust lanes inside spiral arms obscure bright spiral arms, and hence dust lanes are visible on the far side of the disk. This interpretation on the near side of the tilted disk was directly connected to the spiral winding direction, whether the spiral arms observed are trailing spiral or leading spiral \citep{Vaucouleurs1959}.

The current commonly adopted observational interpretation is that for disk galaxies, where there is a discernible asymmetric distribution of dust lanes or color gradient along the minor axis, the dust lane dominant side or darker/redder side on the minor axis is the nearer side.

With this interpretation of the near side, \citet{Iye2019} made a study of 146 nearby spiral galaxies and confirmed that all the spiral structures observed in those galaxies, with (1) identified spiral winding sense, S-wise or Z-wise, as projected on the sky, (2) the identified approaching side on the major axis, and (3) the darker/redder side on the minor axis, have trailing spirals rather than leading spirals. This interpretation is also consistent with numerical simulations of differentially rotating disk galaxies that generate trailing spirals rather than leading spirals.

The fact that all spirals are trailing enables robust identification of the sign of the line-of-sight component of the spin vector by identifying its winding direction as S-wise or Z-wise.  \citet{Tadaki2020} and \citet{Iye2021} generated spin catalogs of spiral galaxies to study if there is any parity violation in the large-scale distribution of spin vectors of galaxies.

In this work, we tried to identify the near side of two spiral galaxies in addition to M31 by detecting the difference in the average color of globular clusters (GCs) on both sides of the major axis as the number fraction of reddened globular clusters should be larger on the near side rather than on the far side.

\citet{Schlafly2011} measured colors of Galactic stars with spectra in the Sloan Digital Sky Survey to map galactic extinction. We use their result to correct for the foreground Galactic extinction of extragalactic GCs to isolate additional extinction due to the interstellar dust of the respective host galaxies.

We describe a simple geometrical model to quantify the visibility of reddening asymmetry in section 2.  The application to M31 is given in section 3 and to other spirals in section 4.

\section{A Model of Halo Shadowed by a Disk}
\begin{figure}[h]
 \begin{center}
 \includegraphics[width=8cm]{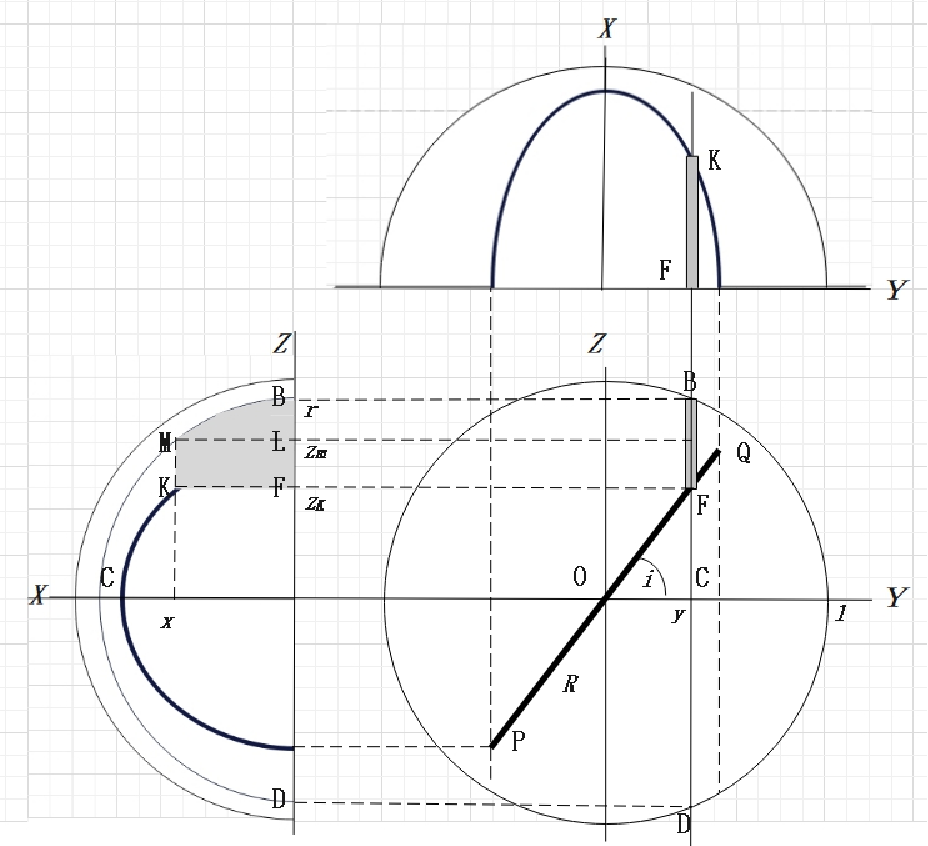} 
 \end{center}
\caption{Geometry of a spherical halo ($R_{h}=1$) containing globular clusters and a tilted ($i$) disk ($R_{d}=R$) shielding the region behind the disk from an observer in the -Z direction. A small circle BCD shows a cross-section of the halo at $Y=y$.}
\label{fig: Shadow}
\end{figure}

Let us assume a simple model, as shown in Figure \ref{fig: Shadow}, where the distribution of globular clusters in a spherical halo of radius $R_{h}=1$ is shielded by a thin disk with interstellar dust of radius $R_{d}=R$ tilted from the normal plane of the line of sight by an inclination $-\pi/2<i<\pi/2$. We take the Z-axis joining the galaxy from an observer at infinity in the $-Z$ direction. The Y-axis is set in the plane orthogonal to the line-of-sight at the galaxy center so that the spin vector of the disk is in the Y-Z plane. The X-axis is orthogonal to the Y-Z plane.

In this configuration, the radius $r$ of the circle section BCD of the halo at $Y=y$ is $r=\sqrt{1-y^{2}}$. The height of the disk at $Y=y$ is given by $Z=z=y\tan{i}$. The two ends (one of which is at K) of the intersection of the disk and the plane $Y=y$ are at $X=\pm x= \pm \sqrt{R^{2}-y^{2}-y^{2}\tan{i}^{2}}$.  

\begin{figure}
 \begin{center}
\includegraphics[width=8cm]{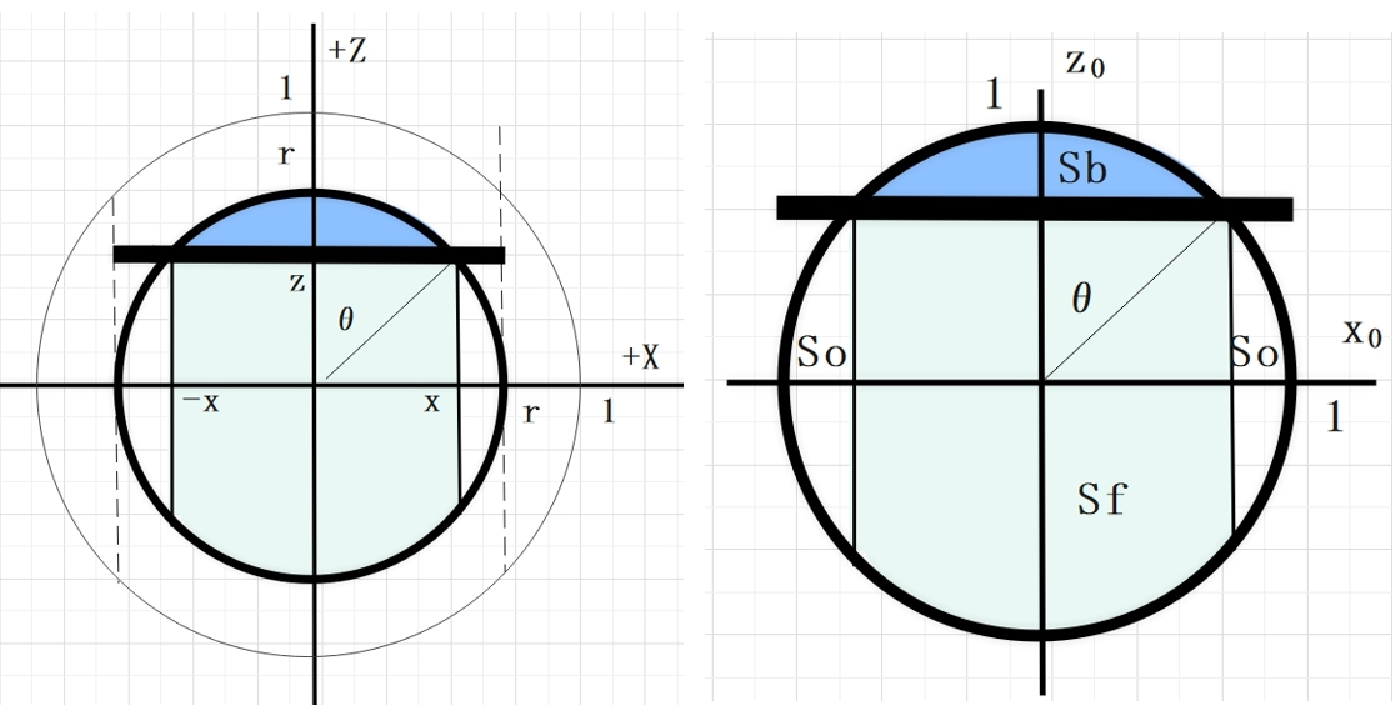}
 \end{center}
 \caption{(Left) Circle cross-section  (thick circle) of a spherical halo (thin circle) and cross-section of the disk (thick horizontal bar) at $Y=y$ plane. (Right) The same figure is shown in coordinates normalized by $r$ to be applied in equations. Areas Sb, Sf, and So show those behind, in front of, and out of disk from an observer at -Z direction for a unit circle.} 
 \label{fig: Shadow0}
 \end{figure}

Figure \ref{fig: Shadow0} shows an example of the X-Z cross-section of the halo at $Y=y$. The disk (thick horizontal line) divides this cross-section into regions behind (Sb), in front of (Sf), and outside of (So) the disk. We denote the areas behind, in front of, and outside the disk, $Sb$, $Sf$, and $So$, respectively, as observed by an observer in the -Z direction.

 By introducing normalized coordinates for the circle cross-section at $Y=y$
 \begin{eqnarray}
 x_{0}=x/r=\frac{\sqrt{R^{2}-y^{2}-y^{2}\tan^{2}{i}} }{\sqrt{1-y^{2}}}
 \end{eqnarray} and
 \begin{eqnarray}
 z_{0}=z/r=\frac{y\tan{i}}{\sqrt{1-y^{2}}}
 \end{eqnarray}
\noindent one can reduce our problem of evaluating the geometrical region affected by disk extinction to finding the fraction $f(z_{0}(y))$ of area Sb to the total area $\pi$ of a unit circle.

In the right panel of Figure \ref{fig: Shadow0}, the fraction $f(z(y))$ of the shaded area of a unit circle, cut by a line segment with endpoints $(-x,z)$ and $(x,z)$, where $x^{2}+z^{2}=1$, is given by $f(z)=\theta-xz$, where $\theta=\arctan{(x/z)}=\arccos{z}$.

Eliminating $x$, we have,
\begin{eqnarray}
f(z) = \cases{ \arccos{z} - z\sqrt{1-z^{2}} & $|z| \le 1$ \cr
                                 0 & $|z| > 1$ \cr}
 \end{eqnarray}

 \begin{figure}[h]
 \begin{center}
\includegraphics[width=8cm]{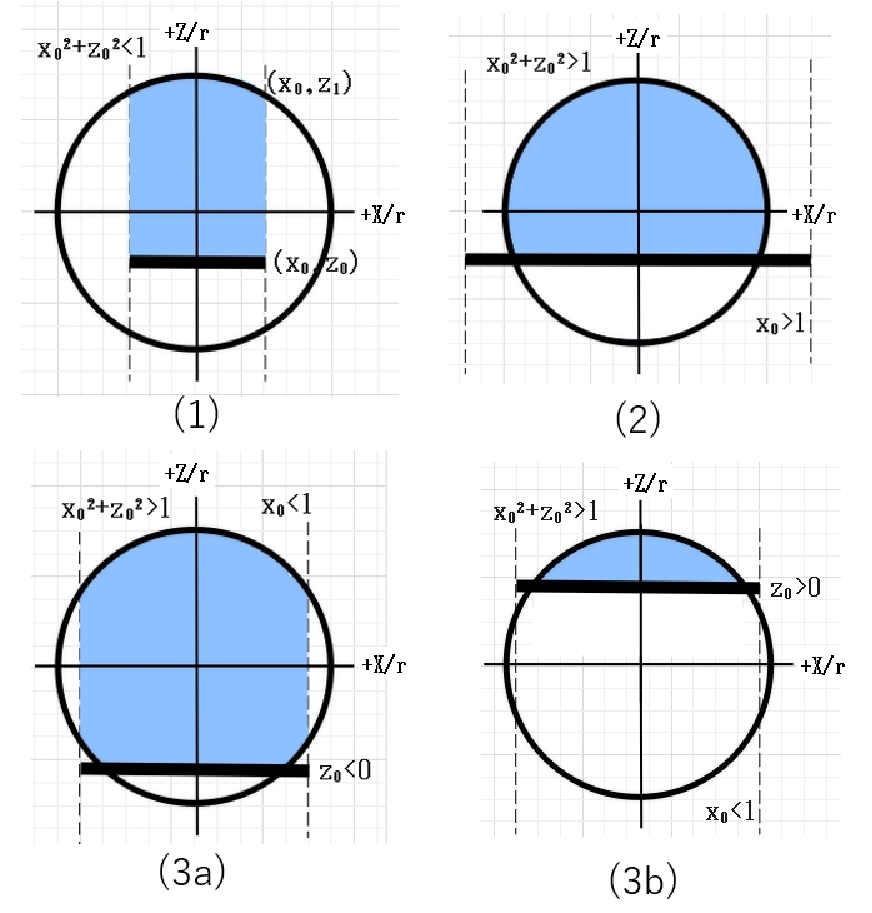}
 \end{center}
 \caption{Four cases of disk shadowing the halo region. } 
 \label{fig: Shadow4Cases}
 \end{figure}

Using equation (1), we can derive an expression for the shadowed area $Sb(y)$ for cases separately with $x_{0}^{2}+z_{0}^{2} \le 1$ and $x_{0}^{2}+z_{0}^{2}>1$.

(1) For a circle section with $\sqrt{ x_{0}^{2}+z_{0}^{2}} \le 1$, as shown in the panel (1) of Figure \ref{fig: Shadow4Cases} for $i<0$,
\begin{eqnarray}
 S_{b}(y) = \cases{ \{f(z_{1})+2x_{0}(z_{1}-z_{0})\}(1-y^{2}) & $0 \le |y| \le |R\cos{i}|$ \cr
                                                                         0 & $ |y|>|R\cos{i}|$ \cr}
                                                                         \end{eqnarray}
\noindent where $z_{1}=\sqrt{1-x_{0}^{2}}$. 

For a circle section with $\sqrt{ x_{0}^{2}+z_{0}^{2}}>1$, there are three cases.

(2)  If $x_{0} \ge 1$, as shown in the panel (2) of Figure \ref{fig: Shadow4Cases} for $i<0$, $S(y)$ is for the entire range $-1 \le y \le 1$
\begin{eqnarray}
S_{b}(y)=\cases{ \{\pi-f(-z_{0})\}(1-y^{2}) & for $z_{0}<0$ \cr 
                    f(z_{0})(1-y^{2}) & for $z_{0}\ge0$ \cr}
\end{eqnarray}

(3a) If $z_{0}<0$, as shown in the panel (3a) of Figure \ref{fig: Shadow4Cases} for $i<0$, 
\begin{eqnarray}
S_{b}(y)=\cases{ \{\pi-f(-z_{0})-2 f(x_{0})\} (1-y^{2}) & $0 \le |y| \le |R\cos{i}|$ \cr
    0 & $|y|>|R\cos{i}|$ \cr}
\end{eqnarray}

(3b) If $z_{0} \ge 0$, as shown in panel (3b) of Figure \ref{fig: Shadow4Cases} for $i>0$, 
\begin{eqnarray}
S_{b}(y)=\cases{ f(z_{0}) (1-y^{2})  & $0 \le |y| \le |R\cos{i}|$ \cr
                0 & $|y|>|R\cos{i}|$ \cr}
\end{eqnarray}

The halo volume behind the disk, $V_{b}$, is obtained by integrating $\pi(1-y^{2})S_{b}(y)$ over $[-R\cos{i}, R\cos{i}]$.
\begin{eqnarray}
V_{b} &=& \pi\int_{-R\cos{i}}^{R\cos{i}} (1-y^{2}) S_{b}(y)dy,
\end{eqnarray}
except for case (2), where the integration is over the range [-1,1].

The volume in front of the disk is given similarly by
\begin{eqnarray}
V_{f} &=& \pi\int_{-R\cos{i}}^{Rcos{i}} (1-y^{2})(1-S_{b}(y))dy,
\end{eqnarray}
except for case (2), where the integration is over the range [-1,1].

The asymmetry fraction $A(i, R)$ can be defined as follows:
\begin{eqnarray}
A(i,R) =  |V_{b}-V_{f}|/(V_{b}+V_{f})
\end{eqnarray} 

Note that $V_{b}+V_{f}  = 4\pi/3$ only for case (2). There are unshaded regions of the halo outside the edge of the shading disk for cases (1), (3a), and (3b), whose volume $V_{o}$ is given by
\begin{eqnarray}
V_{o}= 2\pi \int_{R\cos{i}}^{1} (1-y^{2}) dy = (\frac{4}{3}- 2R\cos{i} +\frac{2R^{3}\cos{i}^{3}}{3})\pi
\end{eqnarray}
for $R < 1$ and $V_{o}=0$ for $R\ge1$.

One can define a volume fraction $B(i, R)$ of the halo region where the detection of the reddening effect of the disk is concerned,
\begin{eqnarray}
B(i, R) = (V_{b}+V_{f})/((V_{b}+V_{f}+V_{o}).
\end{eqnarray}

These two fractions are both in the range (0,1) and the detection of the interstellar reddening effect is easier for a spiral galaxy with large $A(i, R)$ and $B(i, R)$.

The number of reddened GCs is expected to be proportional to $B(i, R)$ and the statistical significance proportional to its square root. 
We can define a geometrical asymmetry visibility parameter $G(i, R)$ for a spiral galaxy by combining these two fractions
\begin{eqnarray}
G(i, R) = A(i, R)\sqrt{B(i, R)}.
\end{eqnarray}

$G(i, R)$ satisfies $0 \le G(i, R)<1$, and a larger $G(i, R)$ means a favorable geometry for detecting the differential reddening of the GC colors.

\begin{figure}[h]
 \begin{center}
\includegraphics[width=5cm]{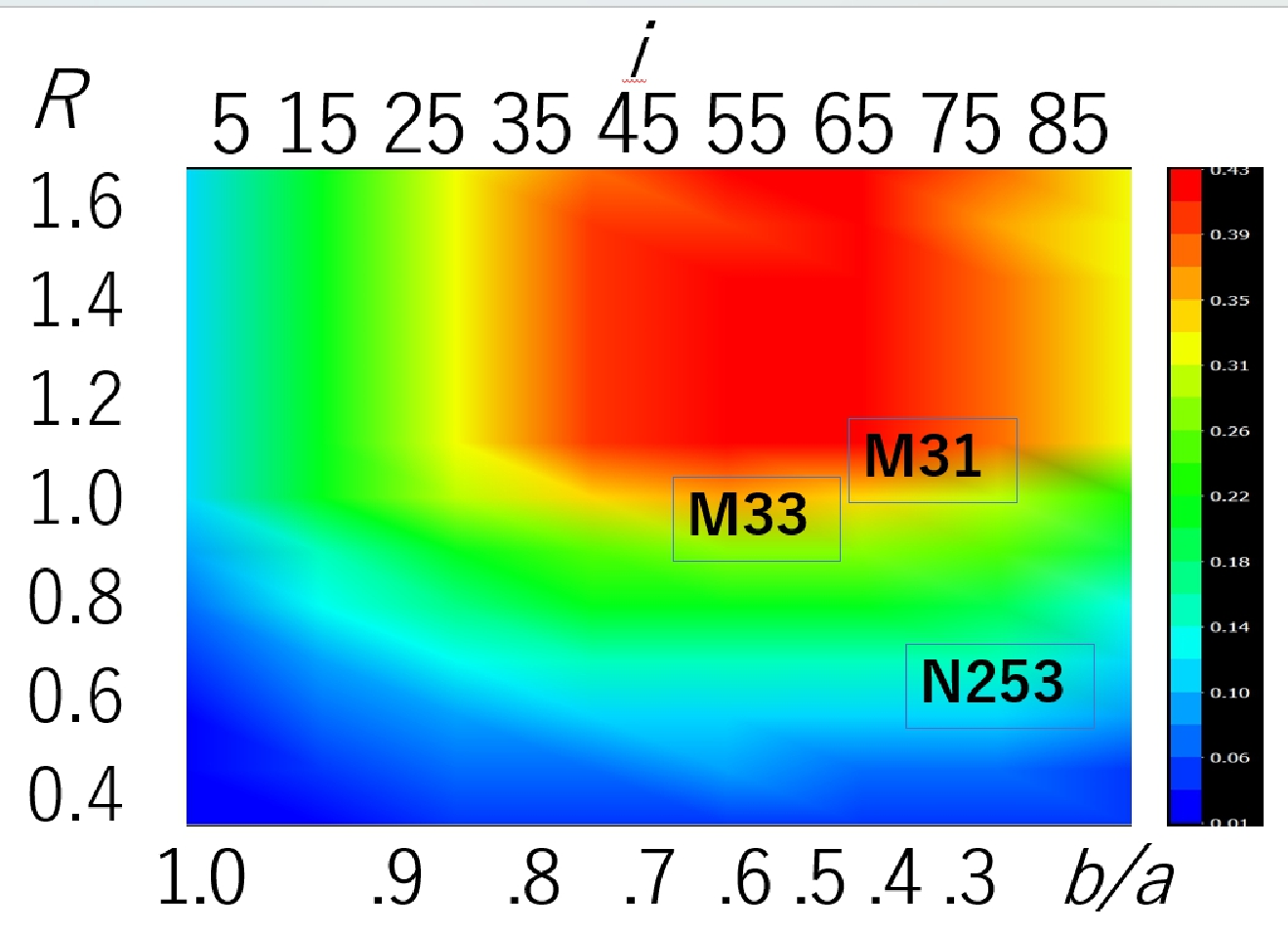}
 \end{center}
 \caption{Geometrical asymmetry visibility $G(i, R)$ as a function of disk size relative to halo size $R=R_{d}/R_{h}$ and inclination of the disk $i=\arccos(b/a)$ from the celestial plane. The locations of the three spiral galaxies studied are shown in this plane.} 
 \label{fig: Visibility}
 \end{figure}

Figure \ref{fig: Visibility} shows the geometrical asymmetry visibility $G(i, R)$ and the parameters for the nearby tilted spirals studied in this paper.  Here, the $R_{h}$ was defined by a radius within which 95\% of the sampled globular clusters belong. $R_{d}=D25/2$ was taken from RC3\citep{Vaucouleurs1991}.  Actual detection depends on the column density and the distribution of interstellar dust and the quality and adopted color bands of the observational photometry of GCs.

\section{Reconfirmation for M31(NGC224)}

\citet{Iye1985} made a photometric study on the distribution of globular clusters (GCs) color for M31 and showed that the identification of the near side of the tilted disk could be made from differential reddening along the minor axis of the galactic disk as there are more reddened GCs behind the galactic plane on the near side rather than on the far side (cf. Figure \ref{fig: 
M31GCIllust}).  The verification was made from three independent photographic photometries available at that time for M31, all showing a consistent result that the northwest side of the M31 disk is the near side, which is the dust lane dominant side.

\begin{figure}[h]
 \begin{center}
  \includegraphics[width=8cm]{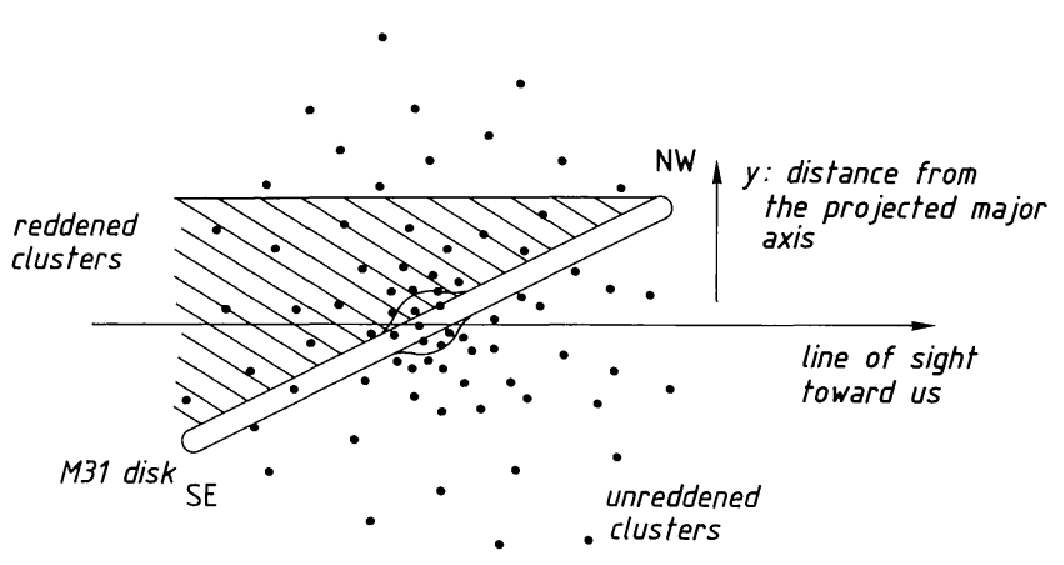} 
 \end{center}
\caption{Illustration of differential reddening of M31 globular clusters. (Figure 1 from \citet{Iye1985})}\label{fig: M31GCIllust}
\end{figure}

\citet{Barmby2000} published a catalog of CCD photometric and spectroscopic data on M31 GCs. Their catalog contains 268 cluster candidates with optical photometry in four or more filters of which objects with $B-V <0.55$ are not likely true globular clusters. Among those objects with $B-V > 0.55$, their catalog provides an estimate of metallicity [Fe/H] for 130 objects from spectroscopic observation and the estimated interstellar reddening according to an empirical relation 
$$E(B-V) = (B-V)-0.159[Fe/H]-0.92,$$
using Table 7 of \citet{Barmby2000}.

\begin{figure}[h]
 \begin{center}
  \includegraphics[width=4cm]{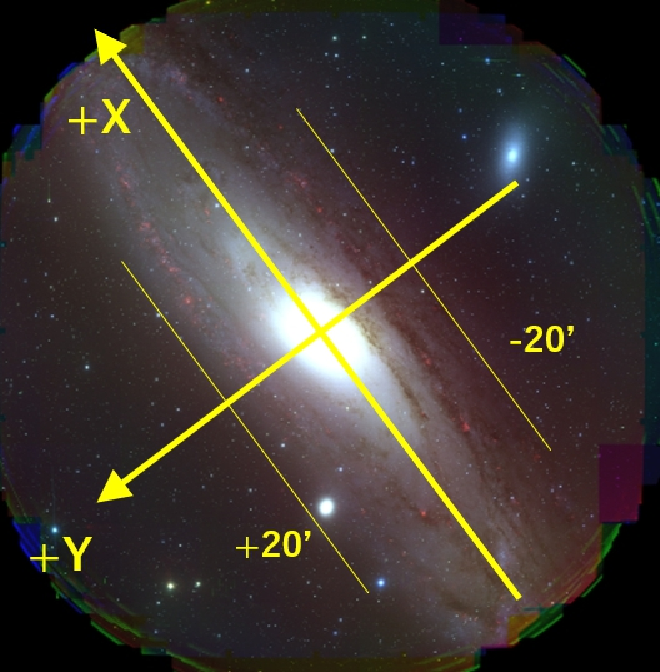} 
 \end{center}
\caption{Coordinate geometry of M31. 
\\Image from \url{https://subarutelescope.org/jp/gallery/pressrelease/024/04/03/3382.html}}\label{fig: M31}
\end{figure}

Figure \ref{fig: M31} shows the coordinate geometry of M31. We take the +Y axis in the direction rotated 90$^\circ$ counterclockwise from the +X axis, which is the direction of the receding side on the major axis. We use this convention for other spirals as well.  Note that these coordinates are not identical to those adopted in \citet{Iye1985}.

\begin{figure}[h]
 \begin{center}
  \includegraphics[width=8cm]{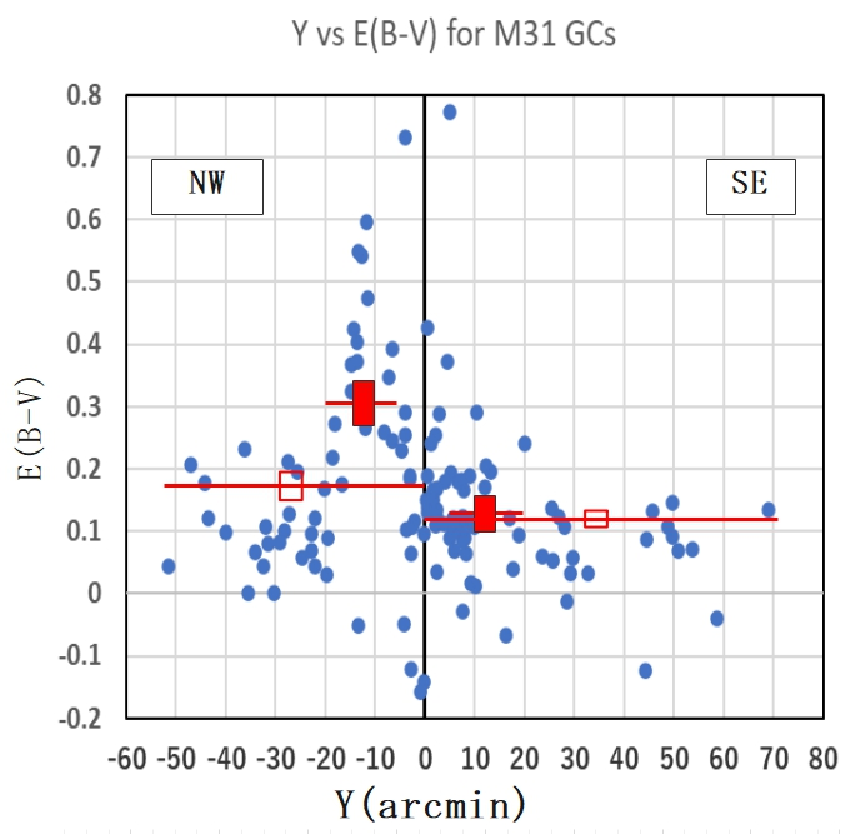} 
 \end{center}
\caption{$E(B-V)$ distribution of globular clusters (blue dots) as a function of $Y$, the distance from the major axis of the M31 disk in arcmin units. The average reddening values and associated standard errors are given in red-filled squares and red open squares over the sampled ranges, $5'<|Y|<20'$, and the entire $Y$ range is shown in horizontal red lines. The figure shows enhanced reddening for GCs behind the disk on the northwest side at domains $-20' < Y <-5'$.}\label{fig: M31GC_E(B-V)}
\end{figure}

Figure \ref{fig: M31GC_E(B-V)} shows the distribution of $E(B-V)$ of globular clusters as a function of position $Y$ in arcmin along the minor axis of M31. %A typical photometric error of $E(B-V)$ measurement is quoted at 0.06 mag \citet{Barmby2000}.  
The root mean square $e_{B-V}$ of individual measurement errors of the sample is 0.03 mag. 

\begin{table}[h]
  \begin{tabular}{cccccc}
      \hline
     color&Region($'$)&\#&Mean&StdErr&$p$ \\ 
      \hline
     B-V&$Y>0$&72&0.858&0.019&\\
     B-V&$Y<0$&58&0.901 &0.027&0.093 \\
     E(B-V)&$Y>0$&72&0.121&0.014&\\
     E(B-V)&$Y<0$&58&0.173&0.024&0.031\\ 
     B-V&$20>Y>5$&32&0.854&0.030& \\
     B-V&$-20<Y<-5$&20&1.018&0.047 &0.0031 \\ 
     E(B-V)&$20>Y>5$&32&0.126&0.024&\\
     E(B-V)&$-20<Y<-5$&30&0.304&0.039&0.00022\\
      \hline
    \end{tabular}\label{tab: M31GC}
\caption{Color statistics of M31 GCs.  The associated standard errors, StdErr, are the estimated errors of the average color $B-V$ and reddening $E(B-V)$. The Welch t-test $p$-values are given for each pair. } 
\end{table}

There is no physical reason to expect a difference in the intrinsic colors of GCs on both sides of the major axis. However, interstellar reddening of the M31 disk would have different effects on both sides, and the standard deviations of the observed color of GCs could be different.

Welch t-test $p$-values for the statistical probability of finding larger mean values of $B-V$ or $E(B-V)$ on the expected near region of the major axis $-Y_{max} < Y < Y_{min}$ than for the far region $Y_{min}  < Y < Y_{max}$ by chance were evaluated for various choices of $Y_{max}$ and $Y_{min}$ at 5$'$ steps. We observe a clear $E(B-V)$ enhancement in the region $-20'<Y<-5'$.  A similar enhancement is also seen in $B-V$ there.

 Table 1 summarizes the results for the entire sample and the sample with the lowest $p$-values. The GCs in $-20'<Y<-5'$ are redder than those in $5'<Y<20'$ by 0.16 mag and more reddened by 0.18 mag. These differences can occur to chance coincidence probability $p$-values 0.0031 or 0.00022, respectively. The $p$-value increases to 0.093 and 0.031, respectively, if we take all the GCs on both sides of the major axis, probably because of the increased population of outer GCs that are not suffering interstellar reddening and inner ($|Y|<5'$) GCs for which $S_{b}$ is not much different from 0.5. 

These $p$-values were calculated using the observed standard deviation of GC colors, including the intrinsic color scatter and reddening effect.  The actual significance of the mean color difference should be compared using only the standard deviation of the inherent color. Therefore, the relevant $p$-values could be smaller than the values in Table \ref{tab: M31GC}.

With the inclination $i=71^{\circ}$ and the disk size normalized to the GC halo size $R=R_{d}/R_{h}=1.1$, the geometrical asymmetry visibility defined by equation (7) is $G=0.38$ for M31.

The presence of anisotropic reddening of the globular cluster colors of M31, as reported in \citet{Iye1985} and supported by \citet{Barmby2000} is reconfirmed and provides another basis for using GCs to judge the near side of the tilted disk of galaxies in the following section.

\section{Application to Other Nearby Spiral Galaxies}

\subsection{NGC253}

\citet{Cant2018} made \it{ugri} \rm photometry of 82 bona fide GCs, 155 uncertain GC candidates, and 110 unlikely GCs (most likely background galaxies) of NGC253. We use their $g-r$ color data, which have smaller photometric errors than their $u-g$ color data.  They applied a constant Galactic extinction to their image frames. We applied Galactic extinction individually for objects based on \citet{Schlafly2011}. 

\begin{figure}[h]
 \begin{center}
  \includegraphics[width=4cm]{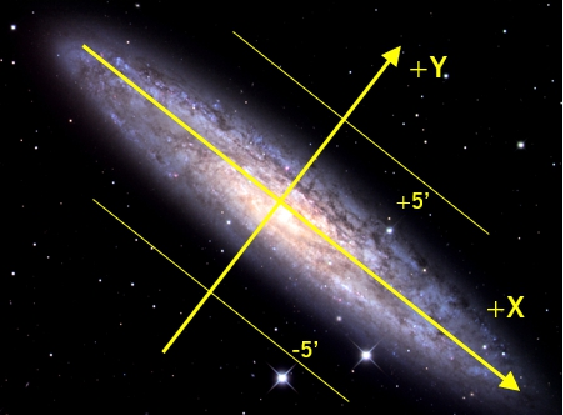} 
 \end{center}
\caption{Geometry of NGC253.\\
Image from \url{https://ps1images.stsci.edu/cgi-bin/ps1cutouts}}\label{fig: N253}
\end{figure}

Figure \ref{fig: N253} shows that the dark lanes are more obvious on the northwest side of NGC253 as cataloged in \citet{Iye2019}.  The axis ratio is $R25=log(a/b)=0.61$ \citep{Vaucouleurs1991}, where $a$ and $b$ are the semimajor and semiminor radii of the galactic disk. The NGC253 disk is slightly more inclined than that of M31 with $R25=0.49$.

\begin{figure}[h]
 \begin{center}
  \includegraphics[width=8cm]{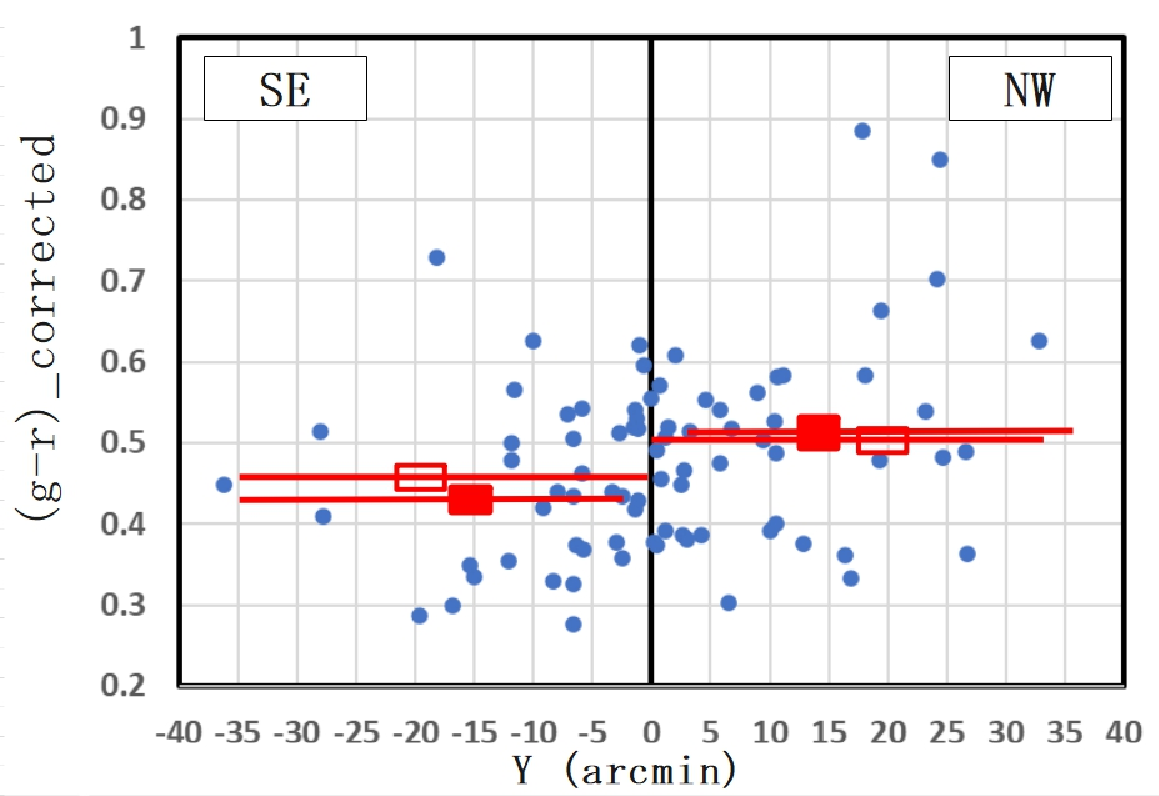} 
 \end{center}
\caption{$Y$ vs $g-r$ distribution of 81 bona fide GCs in NGC253.   The height of the open and closed squares correspond to the mean error of the evaluated average color over the entire region and region $|Y|>2.'5$, respectively.}\label{fig: N253_g-r}
\end{figure}

\begin{table}[h]
   \begin{tabular}{cccccc}
      \hline
     $objects$&Region($'$)&\#&Mean&StdErr&$p$ \\ 
      \hline
     GC*&$Y<0$&39&0.528&0.018&\\
      GC*&$Y>0$&42&0.569&0.019&0.059\\      
      GC&$Y<0$&39&0.456&0.017&\\
     GC&$Y>0$&42&0.504&0.019&0.033\\
       GC&$Y<-2.5$&30&0.435&0.019&\\
      GC&$2.5<Y$&32&0.509&0.024&0.009\\
    galaxies&$Y>0$&60&0.678&0.029&\\
    galaxies&$Y<0$&50&0.673&0.030&0.449\\
      \hline
    \end{tabular}
    \label{tab: NGC253}
\caption{$g-r$ color statistics for NGC253 bona fide GCs and background galaxies. Colors for GC* are uncorrected for foreground Galactic extinction. }
\end{table}

Figure \ref{fig: N253_g-r} shows the distribution of $g-r$ color as a function of $Y$ for bona fide GCs of NGC253, each corrected for the foreground galactic extinction.  Photometric errors in measuring their color depend on the GC brightness.  The root mean square $e_{g-r}$ of the quoted photometric errors for each GC of this sample is 0.01.  Table 2 summarizes the average color on both sides from the major axis. The GCs on the northwest side are redder on average by 0.05 mag than those on the southeast side, indicating that the northwest side is the near side of the tilted disk of NGC253.  The Welch t-test $p$-value evaluations were performed for $2.'5$ step choices of $Y_{max}$ and $Y_{min}$. The lowest $p$-value was obtained for $|Y_{min}|=2.'5$ and no limit for $Y_{max}$.

The Welch-test probability of finding by chance the observed asymmetry in the $g-r$ color for all 81 bona fide GCs is $p=0.033$. The outer 62 GCs at $|Y|>2.5'$ show more conspicuous asymmetry $p=0.009$ with a color difference of 0.07 mag. The statistical significance for asymmetry is slightly decreased to $p=0.059$ for the entire GC sample without correction for the foreground Galactic extinction.

Note that the halo radius $R_{h}=27'$ of the sampled GCs in NGC253 is significantly larger than the disk size $R_{d}=13.8'$, $R=R_{d}/R_{h}=0.51$, compared to $R=1.09$ for M31. This makes the geometrical asymmetry visibility for NGC253 $G=0.06$, approximately 1/6 of $G=0.38$ for M31, despite its large inclination $i=76^{\circ}$. Nevertheless, differential reddening was perceptible for NGC253 because of the large number of GCs available with high-precision photometry. 
%It seems that the NGC253 GC sample contains many unreddened outside GCs.

Note also that Table 2 shows that the sample of background galaxies does not show any difference in their mean color on either side of the major axis of NGC253.
The standard deviation of the $g-r$ color for the background galaxies is 0.22 mag, which is larger than that of the sampled GCs, 0.12 mag.

%%%%%%%%%%%%%%%%%%%%%%%%%%%%%%%%%%%%%%%

\subsection{M33(NGC598)}

\citet{Fan2014} reported $UBVRI$ photometry, ages, metallicities, and masses of 708 star clusters and cluster candidates of NGC598 (M33). We used $B-V$ and the age of their catalog to study the color and extinction dependence on $Y$ by isolating 105 clusters with an evaluated age older than 2 Gyr. 

\begin{figure}[h]
 \begin{center}
  \includegraphics[width=4cm]{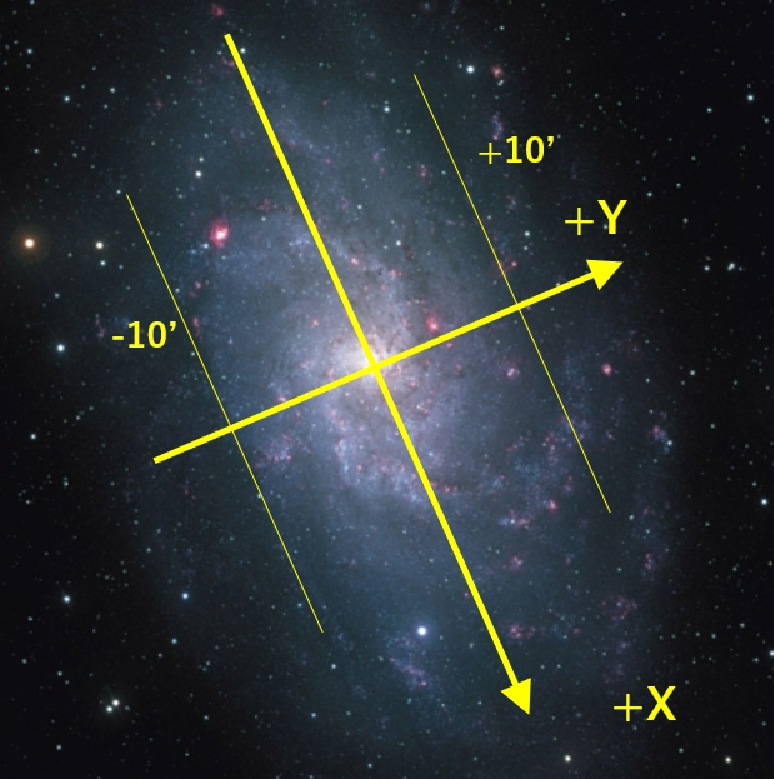} 
 \end{center}
\caption{Geometry of M33. \\
Image from \url{https://www.nao.ac.jp/en/contents/gallery/2017/20170307-subaru-full.jpg}}\label{fig: M33}
\end{figure}

\begin{figure}[h]
 \begin{center}
  \includegraphics[width=5cm]{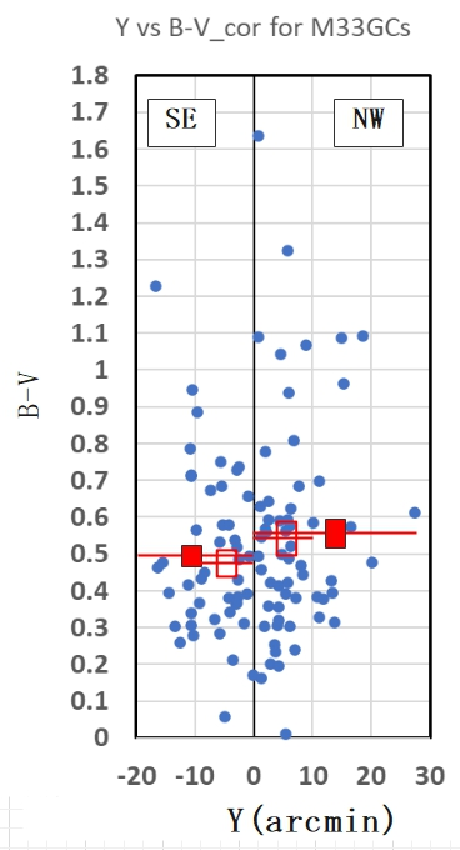} 
 \end{center}
\caption{Distribution of $B-V$ corrected for Galactic extinction as a function of $Y$ for 105 GCs older than 2Gyr in M33.}\label{fig: M33cor}
\end{figure}

 Figure \ref{fig: M33} shows the geometry of M33. Figure \ref{fig: M33cor} shows the distribution of $B-V$ color corrected for Galactic extinction as a function of the position $Y$ from the major axis. 
 
 Table \ref{tab: M33B-V} indicates naively that the $B-V$ color of GCs at $Y>0$ is redder than those at $Y<0$ by 0.06 mag but at a significance level of $p=0.118$. The significance level increases for their raw color, uncorrected for the foreground Galactic extinction, to $p=0.061$. The significance increases slightly for a sample limited to GCs at $|Y|<10$ to $p=0.110$ and $p=0.049$, corrected and uncorrected for individual foreground Galactic extinction, respectively.  The measured color difference is 0.07 mag for the sample at $|Y|<10$. 
 
 The root mean square of the color measurement errors $e_{B-V}$ of M33 is 0.04 mag, larger than 0.03 mag and 0.01 mag for M31 and NGC253, respectively.  This makes the detection of differential reddening rather marginal.  Although the geometric asymmetry visibility parameter $G=0.35$ is similar to the value $G=0.38$ for M31, differential reddening was not securely observed for M33.  The small color difference in M33 is partly due to its smaller inclination of the disk, but it is more likely that M33 has a smaller column density of dust along the line of sight than M31. 

\begin{table}[h]
   \begin{tabular}{cccccc}
      \hline
     Objects&Region($'$)&n&Mean&StdErr&$p$ \\ 
      \hline
    GC*&$Y<0$&45&0.585&0.033&-\\
    GC*&$Y>0$&60&0.664&0.038&0.061\\
    GC*&$-10<Y<0$&31&0.568&0.032&-\\
    GC*&$10>Y>0$&45&0.662&0.046&0.049\\
    GC&$Y<0$&45&0.495&0.033&-\\
    GC&$Y>0$&60&0.556&0.039&0.118\\ 
    GC&$-10<Y<0$&31&0.473&0.033&-\\
    GC&$10>Y>0$&45&0.544&0.047&0.110\\
          \hline
    \end{tabular}\label{tab:M33B-V}
\caption{$B-V$ color asymmetry of M33 GC system.}
\label{tab: M33B-V}
\end{table}
%%%%%%%%%%%%%%%%%%%%%%%%%%%%

\begin{table*}[t]
   \begin{tabular}{cccccccccc}
      \hline
     objects&$L$&$B$&$a(=R_{d})$&$b$&$R_{h}$&PA+&$i$&$R_{d}/R_{h}$&$G(i,R)$\\
    &&&&Regions&Color&$e_{color}$&$n$&$p$&Near Side\\ \hline
     unit&deg&deg&arcmin&arcmin&arcmin&deg&deg&\\
%\\
      \hline
     M31&121.17&-21.57&95.27&30.83&87&35(NE)&71&1.1&0.38\\
     &&&&$5<|Y|<20$&$B-V$&0.03&52&0.0013&NW\\\hline
     M33&133.61&-31.33&35.40&20.84&36&203(SW)&54&1.0&0.35\\
     &&&&$|Y|<10$&$B-V$&0.04&76&0.110&NW\\\hline
     NGC253&97.36&-87.96&13.77&3.38&27&232(SW)&76&0.5&0.06\\
     &&&&$2.5<|Y|$&$g-r$&0.01&62&0.009&NW\\
      \hline
    \end{tabular}\label{tab: summary}  
\caption{Three spiral galaxies were studied for differential reddening of their GC systems. PA+ is the position angle of the receding side of the major axis measured from the north counterclockwise toward the east. $G(i, R)$ is the geometrical asymmetry visibility defined by equation (7).  The Welch t-test probability $p$ of finding the observed color asymmetry by coincidence is given for each pair.} 
\label{tab: summary}
\end{table*}

\section{Summary}
We present a simple geometrical model to evaluate the effect of differential reddening of a globular cluster system of a spiral galaxy with an inclined disk containing interstellar dust.
Table \ref{tab: summary} summarizes geometric parameters of the sampled regions, the observed color data for examining differential reddening, and the $p$-values of three nearby spiral galaxies. The presence of differential reddening of GCs due to an inclined obscuring disk of interstellar dust is confirmed for spirals with high inclination M31 ($R25=0.49$) and NGC253 ($R25=0.61$), but not in a decisive way for a low inclination spiral M33 ($R25=0.23$).  
The correction of foreground Galactic extinction for individual objects slightly increased the significance of detection for NGC253 but not for M31 and M33.

\begin{ack}
This study was supported by a Grant-in-Aid for Scientific Research (C) JSPS KAKENHI 22K03679.
\end{ack}

  \end{document}